\documentclass[a4paper]{article}
\usepackage{geometry}
\usepackage{amsmath}
\usepackage{amsfonts}
\usepackage{amsmath,amssymb}
\usepackage{graphicx}
\usepackage{color}
\usepackage{indentfirst}
\usepackage{threeparttable}

\renewcommand{\baselinestretch}{1}

\newtheorem{theorem}{Theorem}
\newtheorem{lemma}{Lemma}

\newtheorem{example}{Example}

\newcommand{\ls}[1]
    {\dimen0=\fontdimen6\the\font\lineskip=#1\dimen0
     \advance\lineskip.5\fontdimen5\the\font
     \advance\lineskip-\dimen0
     \lineskiplimit=0.9\lineskip
     \baselineskip=\lineskip
     \advance\baselineskip\dimen0
     \normallineskip\lineskip\normallineskiplimit\lineskiplimit
     \normalbaselineskip\baselineskip
     \ignorespaces}

\begin{document}

\bibliographystyle{abbrv}
\title{Quantum Synchronizable Codes on Sextic Cyclotomy}
\author{Tao Wang$^{\mathrm{a}}$,  Tongjiang Yan$^{\mathrm{a}}$,  Xueting Wang$^{\mathrm{a}}$\\
$^a$ College of Science,
China University of Petroleum,\\
Qingdao 266555,
Shandong, China\\
Email: 1409010215@s.upc.edu.cn; yantoji@163.com; \\
s20090013@s.upc.edu.cn}
\maketitle
\thispagestyle{plain} \setcounter{page}{1}
\begin{abstract}
Quantum synchronizable codes are kinds of quantum error-correcting codes that can not only correct the effects of quantum noise on qubits but also the misalignment in block synchronization. In this paper, the quantum synchronizable codes constructed are CSS quantum error-correcting codes whose synchronization capabilities reach the upper bound. And we use cyclic codes gained by sextic cyclotomic classes to construct two classes of quantum synchronizable codes. Moreover, the quantum synchronizable codes are posses good error-correcting capability towards bit error and phase error, since the cyclic codes we used are optimal or almost optimal.


\end{abstract}

\ls{1.2}
\section{Introduction}\label{intro}
In the last two decades, quantum information theory has made great progress in quantum information and quantum communication, especially in quantum error-correcting codes \cite{2000Quantum,1995Scheme}. Since the properties of quantum state, quantum error-correcting codes have some advantages that classical error-correcting codes do not have. The theory and construction of quantum error-correcting codes have extensively been studied \cite{2002QuantumC,1998Good,2020Quantum,PMID:10062908,Steane1996Multiple}. However, quantum error-correcting codes tend to just focus the simplest Pauli errors $I,\; X,\; Y,\; Z$, which roughly corresponds to additive noise in classical encoding theory.

Moreover, as a type of error in block synchronization, misalignment can also cause catastrophic failure in quantum information transmission. This kind of error occurs because the information processing device misidentifies the boundaries of an information stream. For instance, suppose that the quantum information can be expressed by an ordered sequence of information block and each chunk of information is encoded into a block of consecutive three qubits in a stream of qubit $|q_i\rangle$. If three blocks of information are encoded, we have $9$ ordered qubits $\left(|q_0\rangle|q_1\rangle|q_2\rangle|q_3\rangle|q_4\rangle|q_5\rangle|q_6\rangle|q_7\rangle|q_8\rangle\right)$ in which each of the three blocks $\left(|q_0\rangle|q_1\rangle|q_2\rangle\right)$, $\left(|q_3\rangle|q_4\rangle|q_5\rangle\right)$ and $\left(|q_6\rangle|q_7\rangle|q_8\rangle\right)$ forms an information chunk. Suppose the synchronization system was established at the beginning of information transmission, but synchronization may be lost during the quantum communications or quantum computations. This type of error occurs when the receiver incorrectly locates the boundary of each block of data by a certain number of positions towards the left or right. For example, the receiver wrongly read out $\left(|q_5\rangle|q_6\rangle|q_7\rangle\right)$ instead of the correct information chunk $\left(|q_6\rangle|q_7\rangle|q_8\rangle\right)$. \cite{BOSE1967616,2012Block} give more details.

As the subclass of quantum error-correcting codes, quantum synchronizable codes mainly function is to correct the errors caused by quantum noise and block synchronization in the process of quantum information transmission. In order to ensure information security, it is of great significance to study the construction of quantum synchronizable codes \cite{fujiwara2013algebraic,fujiwara2014quantum}. In 2013, Y. Fujiwara\cite{2012Block} proposed the framework of quantum block synchronization and gave the first example of quantum synchronizable codes with standard quantum error correction capability and a general construction framework. In 2014, Xie Yixuan \cite{2014QuantumXie} proposed to use quadratic residual codes and hypercodes to construct quantum synchronizable codes with Mersenne prime numbers. The simple construction of the quantum synchronizable codes given by them can make its synchronization ability reach the upper limit. It also have been proved that it has good error correction performance. After that, quantum synchronizable codes were further studied in \cite{fujiwara2014quantum,2019QuantumLi,XiaLi2021A,2018Non,2016q}.

In this work, we propose a new method to construct the quantum synchronizable codes from the cyclic codes obtained by using sextic cyclotomic classes over $\mathrm{F}_q$. This paper is arranged as follows. In section $\ref{section2}$, we review some general conclusions of cyclic codes and give some knowledge of sextic cyclotomic classes. In section $\ref{section3}$, we briefly review the construction of quantum synchronizable codes and the process of synchronization recovery for quantum information. Section $\ref{section4}$ construct some dual-containing cyclic codes and their augmented codes. In section $\ref{section5}$, we constructed two classes quantum synchronizable codes and discuss their error correction capabilities. Finally, Concluding remarks are given in section $\ref{section6}$.
\section{Preliminaries}\label{section2}
\subsection{Cyclic codes and dual codes}
Let $\mathrm{F}_q$ be a finite field, where $q$ is a prime power. A linear code $C=[n,k,d]_{q}$ is a $k$-dimensional subspace of the $n$-dimensional vector space $\mathrm{F}_q^n$ such that $\mathrm{min}\{\mathrm{wt}(v) : v\in C,\; v\neq0\}=d$, where $\mathrm{wt}(v)$ is number of nonzero elements in $v$. A linear code $C=[n,k,d]_q$ is optimal (almost optimal) means that the code reaches some bound in \cite{Brouwer98}. A linear code $C=[n,k,d]_{q}$ over $\mathrm{F}_q$ is a cyclic code if any codeword $c=(c_0,c_1,...,c_{n-1})\in C$ implies $(c_{t},c_{t+1},c_{t+2},...,c_{n-t-1})\in C$, where $t$ is a nonnegative integer. If each codeword in $C$ can be identified with $ I(C)=\{c_0+c_1x+c_2x^2+...+c_{n-1}x^{n-1}|(c_0,c_1,c_2,...,c_{n-1})\in C\} $, then $ I(C) $ can be regarded as an ideal of the principal ring $ \mathrm{F}_q[x]/{(x^n-1)} $. Hence the cyclic code $C$ can be generated by the monic polynomial $g(x)$ of degree $n-k$, $g(x)$ is called generator polynomial of $C$. Moreover, $g(x)$ is a monic divisor of $x^n-1$. Let $C=\langle g(x) \rangle$. Then the set of codewords of cyclic code $C$ can be written as $C=\{m(x)g(x)|m(x) \in \mathrm{F}_q[x],\; \mathrm{deg}\left(m(x)\right)<k\}$.

The Euclidean inner product between two codewords $x = (x_0, x_1, \dots, x_{n-1})$ and $y = (y_0, y_1, \dots, y_{n-1})$ is defined by $(x, y)=\sum\limits_{i=0}^{n-1}x_{i}y_{i}$. For a linear code $C$ with length $n$, the code $C^\perp=\{x\in \mathrm{F}(q)^n|(x,c)=0 , \forall c \in C\}$ is called an Euclidean dual code of $C$. The dimension of $C^\perp$ is $\mathrm{dim}(C^\perp)=n-\mathrm{dim}(C)=n-k$. And the dual of a cyclic code is also a cyclic code \cite{2003Fundamentals}.

Let $g(x)$ be the generator polynomial of a cyclic code $C$. It follows that $g(x)|(x^n-1)$. Let $h(x)=\frac{x^n-1}{g(x)}$. Then $h(x)$ is the parity check polynomial of cyclic code $C$. Since the degree of the generator polynomial $ g(x) $ is $ \mathrm{deg}\left(g(x)\right)=n-k $, the degree of parity check polynomial $ h(x) $ is $ \mathrm{deg}\left(h(x)\right)=k $. Then the generator polynomial of $C^\perp$ has the form:
\begin{equation}\label{Cperp}
 g^{\perp}(x)=\tilde{h}(x)=h(0)^{-1}x^kh\left(x^{-1}\right),
\end{equation}
where $\tilde{h}(x)$ is called the reciprocal polynomial of $h(x)$. Let $C^\perp=\langle g^\perp(x) \rangle$.

Let $ C_1=\langle (g_1(x)\rangle $ be a cyclic code with parameters $ [n,k_1]_q $, and $ C_2=\langle g_2(x)\rangle $ be a cyclic code with parameters $ [n,k_2]_q  $, where $ k_1<k_2 $. Then, if $ C_1\subseteq C_2 $, $ C_2 $ is said to be $ C_1 $-containing, and $ C_1 $ is called the subcodes of $ C_2 $, $ C_2 $ is called the supercodes of $ C_1 $. If $ C_2^\perp \subset C_2 $, $C_2$ is called dual-containing. If $ C_1\subseteq C_2 $, then the generator polynomial $ g_2(x) $ can divide any codeword of $C_1 $, i.e., for any $ c(x)\in C_1 $, there must exist a polynomial $ f_c(x) $ in $ \mathrm{F}_q[x] $ such that $ c(x)=f_c(x)g_2(x) $. If $ f(x)=\frac{g_1(x)}{g_2(x)} $, then $ f(x)\in \mathrm{F}_q[x] $ has degree $ k_2-k_1 $.

\subsection{Cyclotomic classes}

Let $n=6f+1$ be an odd prime and $\gamma$ be a fixed primitive element of $\mathrm{F}_n$. Then the sextic cyclotomic classes $C_0^{(6,n)},\;C_1^{(6,n)},\; \cdots , C_{5}^{(6,n)}$ of $\mathrm{F}_n$ by
\begin{equation*}
C_i^{(6,n)}=\{ \gamma^{6j+i}|0 \leq j \leq \frac{n-1}{6}-1\},
\end{equation*}
where $0\leq i\leq 5$. Trivially $C_i^{(6,n)}=\gamma^iC_0^{(6,n)}$. The number of elements in $C_i^{(6,n)}$ denoted by $|C_i^{(6,n)}|$. So for arbitrary integers $i,\;j$ satisfy $0\leq i\neq j \leq 5$, we have $|C_i^{(6,n)}|=|C_j^{(6,n)}| $, and $ \mathrm{F}_n^*=\cup_{i=0}^{5} C_i^{(6,n)} $, $ C_i^{(6,n)}\cap C_j^{(6,n)}=\emptyset$, where $\mathrm{F}_n^*=\mathrm{F}_n \backslash{ \{0\} }$.

\begin{lemma}\label{lemma1}
Let $n=12m+7 $ be an odd prime, where $m$ is a nonnegative integer. Then
\begin{equation}
 (a)\;C_0^{(6,n)}=-C_3^{(6,n)},\quad (b)\;C_1^{(6,n)}=-C_4^{(6,n)},\quad (c)\;C_2^{(6,n)}=-C_5^{(6,n)}.
\end{equation}
\end{lemma}
$\mathbf{Proof.}$
$Case(a)$ \;Let $\gamma$ be a fixed primitive element of $\mathrm{F}_p$. It is readily apparent that $-1=\gamma^{\frac{n-1}{2}}=\gamma^{6m+3}\in C_3^{(6,n)}$. Thus for any $a \in C_3^{(6,n)}$, it has a form of expression $a=\gamma^{6j+3}$ where $j=\{0,1,...,2m\}$. Furthermore, $-a=\gamma^{6m+3}\gamma^{6j+3}=\gamma^{6(m+j+1)}\in C_0^{(6,n)}$. So, $-C_3^{(6,n)}\subseteq C_0^{(6,n)}$. Since $|C_3^{(6,n)}|=|C_0^{(6,n)}|$, we get $-C_3^{(6,n)}= C_0^{(6,n)}$.

Similarly, the other conclusions can be proved like $Case(a)$. $\blacksquare$

From now on, we let $q \in C_0^{(6,n)}$, and $\eta$ be a $n$-$th$ primitive root of unity over $F_{q^{\mathrm{ord}_n(q)}}$, where $\mathrm{ord}_n(q)$ is the multiplicative order of $q$ modulo $n$. It is known that the polynomials
\begin{equation}\label{g_i(x)}
g_i^{(6,n)}(x)=\prod_{j\in C_i^{(6,n)}}(x-\eta^j),
\end{equation}
for $i=\{0,1,2,3,4,5\}$ belong to $\mathrm{F}_q[x]$, due to $q\in C_0^{(6,n)}$. Thus, we can get the factorization of $ x^n-1 $ as
$$ x^n-1 = (x-1)\prod_{i=0}^5 g_i^{(6,n)}(x).$$
\section{Quantum synchronizable codes}\label{section3}
\subsection{Coding scheme}
An $[[n,k]]$ quantum error-correcting code is a coding theory that encodes $k$ logical qubits into $n$ physical qubits. As in classic linear code, $n$ and $k$ are the length and dimension of the code, respectively. Quantum error-correcting codes tend to just correct the effects of bit error and phase error caused by Pauli operators. Quantum synchronizable code $(c_l,c_r)-[[n,k]]$ is a encode scheme that corrects not only bit errors and phase errors but also misalignment up to the left by $c_l$ qubits and up to the right by $c_r$ qubits, where $c_l$ and $c_r$ are nonnegative integers.

We provide here construction of quantum synchronizable codes by applying the method discovered by Fujiwara \cite{2012Block,fujiwara2014quantum}. Let $f(x)\in \mathrm{F}_q[x]$ be a polynomial over $\mathrm{F}_q$ and $f(0)=1$. The order $\mathrm{ord}\left(f(x)\right)$ of the polynomial $f(x)$ is the smallest integer $a$ such that $f(x)|(x^a-1)$ over $\mathrm{F}_q[x]$.

\begin{lemma}\cite{2012Block}\label{Theorem1}
Let $C_1=\langle g_1(x)\rangle$ and $C_2=\langle g_2(x)\rangle$ be two cyclic code of parameters $[n,k_1,d_1]_q$ and $[n,k_2,d_2]_q$ with $k_1>k_2$ respectively such that $C_2\subset C_1$ and $C_2^{\perp}\subseteq C_2$. Define $f(x)$ of degree $k_1-k_2$ to be the quotient of $\frac{g_2(x)}{g_1(x)}$ over $\mathrm{F}_q[x]/x^n-1$. For any pair of nonnegative integers $c_l,\;c_r$ satisfying $c_l+c_r<\mathrm{ord}\left(f(x)\right)$, then exist a quantum synchronizable code $(c_l,c_r)-[[n+c_l+c_r,2k_2-n]]$ that corrects at least up to $\left \lfloor \frac{d_1-1}{2} \right \rfloor$ bit errors and $\left \lfloor \frac{d_2-1}{2} \right \rfloor$ phase errors, and misalignment up to the left by $c_l$ qubits and up to the right by $c_r$ qubits.
\end{lemma}

The quantum synchronizable code described in Lemma \ref{Theorem1} requires a pair of cyclic codes $C_1$, $C_2$ of the same length, and dimension $k_1>k_2>\left \lceil \frac{n}{2} \right \rceil$. A misalignment of $\delta$ qubits can be recovered of $-c_l<\delta <c_r$, where minus means that a misalignment occurs towards the left, and the maximum capability of quantum synchronization error is given by $c_l+c_r<\mathrm{ord}\left(f(x)\right)$. To get a good quantum synchronizable code, cyclic codes with good minimum distance are required, while ensuring $\mathrm{ord}\left(f(x)\right)$ to be as large as possible.

\subsection{Encoding}
Since $\mathrm{dim}\left(C_2\right)=k_2$, $\mathrm{dim}\left(C_2^{\perp}\right)=n-k_2$ and $C_2^{\perp}\subseteq C_2$. Assuming that $c,\;c'\in C_2$ are two codewords of $C_2$. Define an equivalence relation ``$\approx$'' on $C_2$, $c\approx c' \Leftrightarrow c-c' \in C_2^{\perp}$. Moreover, it is not difficult to see that the equivalence class $\bar{c}$ determined by a codeword $c\in C_2$ is equal to $\bar{c}=c+C_2^{\perp}:=\{c+x|x\in C_2^{\perp}\}$. Thus, the coset $\bar{c}$ form a partition of $C_2$, and the dimension of coset is $\mathrm{dim}\left(C_2/C_2^{\perp}\right)=2k_2-n$. Hence, the number of cosets is $q^{2k_2-n}$. Therefore, the quantum synchronizable codes encodes $2k_2-n$ logical qubits. Let the representative of cosets denote $\mathcal{B}=\{b_i(x)|0\leq i \leq q^{2k_2-n}\}$. Then the set
\begin{equation}\label{e2}
V=\{|C_2^{\perp}+b_i(x)+g_1(x)\rangle |b_i(x)\in \mathcal{B} \}
\end{equation}
of $q^{2k_2-n}$ quantum states forms an orthogonal basis of a vector space of dimension $q^{2k_2-n}$, where
$$|C_2^{\perp}+b_i(x)+g_1(x)\rangle=\frac{1}{\sqrt{|C_2^{\perp}|}}\sum_{c_i^{\perp}(x)\in C_2^{\perp}} |c_i^{\perp}(x)+b_i(x)+g_1(x)\rangle.$$

By using the standard encoder for Calderbank-Shor-Steane(CSS) \cite{2020Quantum} codes, we encode an arbitrary $q^{2k_2-n}$-qubit state $|\psi\rangle$ into $n$-qubit state $|\psi\rangle_{enc}=\sum_{i}\alpha_i |v_i\rangle,$ where $v_i \in V$. Moreover, let $c_l$, $c_r$ be nonnegative integers such that $c_l+c_r<\mathrm{ord}\left(f(x)\right)$, ancilla qubits used to attach to the left and right of the original state respectively and then applying $\mathrm{CNOT}$ gate, we final obtain an $n+c_l+c_r$-qubit state $|\Psi\rangle_{enc}=|0\rangle^{\otimes c_l}|\psi\rangle_{enc} |0\rangle^{\otimes{c_r}}=\sum_{i}\alpha_i|l_i,v_i,r_i\rangle$, where $l_i$ and $r_i$ are the last $c_l$ and the first $c_r$ positions of the vector $v_i$, respectively. Hence, $|\Psi\rangle_{enc}$ is ready to transmit.

\subsection{Synchrnization recovery}
Here we introduce the procedures for error correction and synchronization recovery. \cite{2020Quantum} gives more details.

Suppose that the receiver gathered qubits of one block length, that is, consecutive $n+c_l+c_r-1$ qubits, and correct bit errors, phase errors and misalignment if necessary. Define block frame $\mathcal{T}=(t_0,t_1,...,t_{n+c_l+c_r-1})$ of the encoded state. If no misalignment occurred, $\mathcal{T}$ forms a properly aligned block encoded as $|\Psi\rangle_{enc}$. Now, we assume that $\mathcal{T}$ may be misaligned by $\delta$ qubits to the right, where $-c_l<\delta<c_r$. When $\delta$ is negative, it means the misalignment occurs to the left by $|\delta|$ qubits.

Define a window frame $\mathcal{W}=(w_{c_l},w_{c_l+1},...,w_{c_l+n-1})$. The device first focus on consecutive $n$ qubits in the middle of $\mathcal{T}$, then $\mathcal{W}=(t_{c_l},t_{c_l+1},...,t_{c_l+n-1})$. Assume misaligned by $\delta$ occurs, this set of qubits is $\mathcal{W}=(t_{\delta+c_l},t_{\delta+c_l+1},...,$ $t_{\delta+c_l+n-1})$.

Let operator $E$ is the $n+c_l+c_r$-fold tensor product of linear combinations of Pauli errors. Therefore the corrupted encoded state is:
$$E|\Psi\rangle_{enc}=\sum_{i}\alpha_{i}(-1)^{(l_i,v_i,r_i)\cdot \boldsymbol{e_{p_i}}}|(l_i,v_i,r_i)+\boldsymbol{e_{b_i}}\rangle,$$
where $\boldsymbol{e_{b}}$ and $\boldsymbol{e_{p}}$ represent bit and phase error respectively.

The receiver corrects errors in two steps, first corrects bit errors on $\mathcal{W}$ and then detects misalignment. Let $H_{C_1}$ be the $(n-k_1)\times n$ parity-check matrix of $C_1$. As in the standard decoding of CSS code, $E|\Psi\rangle|0\rangle^{\otimes n-k_1} \rightarrow E|\Psi\rangle_{enc}|H_{C_1}\boldsymbol{e_{b_i}}\rangle$, where $H_{C_1}\boldsymbol{e_{b_i}}$ is the error syndrome of the window $\mathcal{W}$. If at most $\left \lfloor \frac{d_1-1}{2} \right \rfloor$ bit errors in $\mathcal{W}$, applying $X$ Pauli operators to the qubits can eliminate all bit errors in $\mathcal{W}$.

The next step is the procedure of synchronization recovery. Since $C_2^{\perp}\subseteq C_2 \subset C_1$, the generator polynomial $g_1(x)$ divides every elements in $V$ given in \ref{e2}, which implies
$$|c_i^{\perp}+b_i(x)+g_1(x)\rangle=|v_1(x)f(x)g_1(x)+v_2(x)f(x)g_1(x)+g_1(x)\rangle,$$
for some polynomial $v_1(x)$ and $v_2(x)$ of degree less than $k_2$. Because of the misalignment and window farm $\mathcal{W}$ contains a cyclic shift coefficient vectors of the correct polynomials. Thus, if a misalignment of $\delta$ qubits occurs, the corrupted state has the form $|x^{\delta}(c_i^{\perp}+b_i(x)+g_1(x))\rangle$. So the quotient of $\mathcal{W}$ divided by $g_1(x)$ is $x^{\delta}(v_1(x)f(x)+v_2(x)f(x)+1)$. Dividing this quotient by $f(x)$ gives remainder of $x^{\delta}$:
$$\frac{x^{\delta}(c_i^{\perp}+b_i(x)+g_1(x))}{g_1(x)f(x)}\equiv x^{\delta} \bmod f(x).$$
Thus, if $c_l+c_r<\mathrm{ord}\left(f(x)\right)$, the synchronization error $\delta$ is uniquely determined.
From now on, we know how many qubits the original frame $\mathcal{T}$ is misaligned. By cyclicly shifting $\mathcal{T}$ to recover the correct block of qubits. And any bit and phase errors can be correct as the standard CSS code does.

\section{Cyclic codes from the sextic cyclotomic classes}\label{section4}
In this section, we gave some dual-containing cyclic codes obtained by the sextic cyclotomic classes. Then we discussed their minimum distances. And give some optimal cyclic code finally.

\subsection{Dual-containing Codes}
Let $ C_i $ be the cyclic code by generated $ g_i^{(6,n)}(x) $ and $ \bar{C}_i $ be the cyclic code by generated $ (x-1)g_i^{(6,n)}(x)g_{i+1}^{(6,n)}(x)g_{i+2}^{(6,n)}(x)g_{i+4}^{(6,n)}(x)g_{i+5}^{(6,n)}(x) $, for any $ i\in \{0,1,2,3,4,5\}$.
\begin{lemma}\label{lemma2}
Let $ n=12m+7 $ be an odd prime, where $m$ is a nonnegative integer. Then we have the following conclusions of cyclic codes $ C_i $ and $ \bar {C}_i $:
\begin{equation}
 (a)\;C_i^\perp = \bar {C}_i, \qquad (b)\;C_i^\perp \subset C_i
\end{equation}
\end{lemma}
$\mathbf{Proof.}$
By $(\ref{Cperp})$, the reciprocal polynomial of $g_i^{(6,n)}(x)$ is
$$\tilde{g}_i^{(6,n)}(x)=\left(g_i^{(6,n)}(0)\right)^{-1}x^{\mathrm{deg}\left(g_i^{(6,n)}\right)}g_i^{(6,n)}(x^{-1}),$$
where $g_i^{(6,n)}(x)$ is defined in $(\ref{g_i(x)})$. It follows that
$$g_i^{(6,n)}(x)=(x-\eta^{e_{i_1}})(x-\eta^{e_{i_2}})\cdots (x-\eta^{e_{i_{2m+1}}}),$$
where $C_i^{(6,n)}=\{e_{i_j}|1\leq j\leq2m+1\}$. Then
\begin{equation*}
\begin{split}
\tilde{g}_i^{(6,n)}(x)=&(-\eta^{e_{i_1}})^{-1}(-\eta^{e_{i_2}})^{-1}\cdots(-\eta^{e_{i_{2m+1}}})^{-1}x^{2m+1}(x^{-1}-\eta^{e_{i_1}})(x^{-1}-\eta^{e_{i_2}})\cdots (x^{-1}-\eta^{e_{i_{2m+1}}}) \\
=&(x-\eta^{-e_{i_1}})(x-\eta^{-e_{i_2}})\cdots (x-\eta^{-e_{i_{2m+1}}}).
\end{split}
\end{equation*}
According to Lemma $\ref{lemma1}$, it is readily seen that
\begin{equation}\label{ii3}
\tilde{g}_i^{(6,n)}=g_{i+3}^{(6,n)}(x),
\end{equation}
for any $i=\{0,1,2,3,4,5\}$. The parity check polynomial of $C_i$ is
$$h(x)=\left(\frac{x^n-1}{g_i^{(6,n)}(x)}\right)=(x-1)\prod_{{j\in F_n^{*}\backslash  C_i^{(6,n)}} }(x-\eta^{j}).$$
And by $(\ref{Cperp})$ and $(\ref{ii3})$,
$$\tilde{h}(x) = (x-1)g_i^{(6,n)}(x)g_{i+1}^{(6,n)}(x)g_{i+2}^{(6,n)}(x)g_{i+4}^{(6,n)}(x)g_{i+5}^{(6,n)}(x). $$
Then $C^{\perp}=\langle\tilde{h}(x) \rangle$. According to the definition of $\bar{C}_i$, we get $ C_i^{\perp} = \bar {C}_i $. Moreover, due to $g_i^{(6,n)}(x)| \tilde{h}(x)$, we obtain $ C_i^\perp \subset C_i $. $\blacksquare$

In addition, let $ D_i $ be the cyclic code by generated $ g_{i}^{(6,n)}(x)g_{i+1}^{(6,n)}(x)g_{i+2}^{(6,n)} $, and $ \bar{D}_i $ be the cyclic code with generator polynomial $ (x-1)g_{i+3}^{(6,n)}(x)g_{i+4}^{(6,n)}(x)$ $g_{i+5}^{(6,n)}(x) $ for any $ i\in\{0,1,2,3,4,5\} $. We have the following results, obviously.
\begin{lemma}\label{lemma3}
Let $ n=12m+7 $ be an odd prime, where $m$ is a nonnegative integer. Then the conclusions of cyclic codes $ D_i $ and $ \bar {D}_i $ are
\begin{equation}\label{lemma8}
 (a)\;D_i^\perp = \bar {D}_i,\qquad (b)\;D_i^\perp \subset D_i.
\end{equation}
\end{lemma}
$\mathbf{Proof.}$
The proof is straightforward from the Lemma \ref{lemma2}. $\blacksquare$

\begin{example}
Let $ n = 12m+7 $ and $ q  \in C_0^{(6,n)} $. Table \ref{biaolall} gives some examples about cyclic codes and their duals, and some them are optimal or almost optimal. All computations have been done by MAGMA \cite{Magma} .
\begin{table}[h]
	\centering
	\caption{Dual-containing cyclic codes $C_i$ and $D_i$}  %
    \label{biaolall}
	\begin{tabular}{lll}
    \hline\noalign{\smallskip}
		Codes $C_i$ or $D_i$ &  Duals $C^{\bot}_i$ or $D^{\bot}_i$ &  Optimal or almost optimal \cite{Grassl:codetables} \\
		\noalign{\smallskip}\hline\noalign{\smallskip}
		 $ C_i=[19,16,3]_7 $&$ C^{\bot}_i=[19,3,15]_7 $& Both optimal \cite{Grassl:codetables}  \\
	
		 $ D_i=[19,13,5]_7 $& $ D^{\bot}_i=[19,6,12]_7 $& $D_i$ almost optimal, $D^{\perp}_i$ optimal \cite{Grassl:codetables}\\
	
         $ C_i=[31,26,3]_2 $& $C^{\bot}_i=[31,5,16]_2 $& Both optimal \cite{Grassl:codetables}\\

		 $ D_i=[31,21,5]_2 $& $ D^{\bot}_i=[31,10,12]_2 $& Both optimal \cite{Grassl:codetables}\\
		\noalign{\smallskip}\hline
	\end{tabular}
\end{table}
\end{example}

\subsection{Augmented cyclic codes}
A cyclic code $C'=\langle g'(x) \rangle$ is called augmented cyclic code of $C=\langle g(x) \rangle$ if $C\subset C'$. It follows that $g'(x)|g(x)$. To obtain the augmented cyclic codes of $C_i$ and $D_i$, we need the concept of cyclotomic cosets. For any nonnegative $s$ and $n$, the cyclotomic cosets $C_{(s,n)}$ of $s$ modulo $n$ over $\mathrm{F}_q$ is the set
\begin{equation}\label{DCC}
C_{(s,n)}=\{sq^{i}\bmod n|i\in \mathbb{N}\}.
\end{equation}
Let $\eta$ be a primitive $n$-$th$ root of unity in some extension field of $\mathrm{F}_q$, where $q\in C_0^{(6,n)}$. Then the unique irreducible minimal polynomial of $\eta^{s}$ over $\mathrm{F}_q[x]$ is $M_s(x)=\prod_{i\in C_{(s,n)}}(x-\eta^i) \in \mathrm{F}_q[x]$.

\begin{theorem}
Let cyclic code $C_i=\langle g_i^{(6,n)}(x) \rangle$ be defined above. If the size of $C_{(1,n)}$ is $|C_{(1,n)}|=\ell$, then the generator polynomial $g_i^{(6,n)}$ can be expressed as the product of $t=\frac{n-1}{6\ell}$ irreducible polynomials of degree $\ell$ over $\mathrm{F}_q[x]$ as follows:
\begin{equation*}
g_i^{(6,n)}=\prod_{j=1}^{t}M_{i_j}(x),
\end{equation*}
where the degree of $M_{i_j}$ is $\mathrm{deg}(M_{i_j}(x))=\ell$, for all $i_j$.
\end{theorem}
$\mathbf{Proof.}$
As $n=12m+7$ is an odd prime and $|C_{(1,n)}|=\ell$, then $|C_{(s,n)}|=\ell$ for any integers $s\in\{1,2,...,n-1\}$. Since $q\in C_{0}^{(6,n)}$, it is known that $C_{i}^{(6,n)}=\cup _{j-i}^t C_{(i_j,n)}$, where $i_1,i_2,...,i_t\in \{1,2,...,n-1\}$ are some integers. Moreover, we have $t=\frac{|C_{i}^{(6,n)}|}{|C_{(s,n)}|}=\frac{n-1}{6\ell}$. $\blacksquare$

\begin{example}
Consider the sextic cyclotomic classes modulo $n=127$
\begin{equation*}
\begin{aligned}
C_0^{(6,127)}=\{&1,47,50 ,64,  87,  25,  32, 107,  76,  16, 117,  38,   8, 122,  19,   4,  61,  73,   2,  94,100\}, \\
C_1^{(6,127)}=\{&6 ,28 , 46,   3,  14,  23,  65,   7,  75,  96,  67, 101,  48,  97, 114,  24, 112,  57,  12,  56,92\}, \\
C_2^{(6,127)}=\{&36,  41,  22,  18,  84,  11,   9,  42,  69,  68,  21,  98,  34,  74 , 49 , 17 , 37,  88,  72,  82,44\}, \\
C_3^{(6,127)}=\{&89, 119,   5, 108, 123,  66,  54, 125,  33,  27, 126,  80,  77,  63,  40, 102,  95,  20,51, 111,  10\}, \\
C_4^{(6,127)}=\{&26 , 79,  30,  13, 103,  15,  70, 115,  71,  35, 121,  99,  81, 124, 113, 104,  62, 120,  52,  31,  60\}, \\
C_5^{(6,127)}=\{&29,  93,  53,  78, 110,  90,  39,  55,  45,  83,  91,  86, 105, 109,  43, 116, 118,  85,58,  59, 106\}.
\end{aligned}
\end{equation*}
Let $\gamma$ be the fixed primitive element of $\mathrm{F}_{127}$. Then $q=\gamma^{6K}\in C_0^{(6,127)}$, where $K=\{0,1,...,20\}$. The order of $q$ modulo $n$ is $|q|=|\gamma^{6K}|=\frac{|\gamma|}{\mathrm{gcd}(6K,|\gamma|)}$. For $K=3$, $\gamma=39$, we have $q=39^{18}= 2$ and $|q|=\frac{126}{\mathrm{gcd}(126,18)}=7$ over $\mathrm{F}_{127}$.By (\ref{DCC}), we have
\begin{equation*}
\begin{aligned}
&C_{(1,127)}=\{1  , 2,   4,   8,  16,  32,  64\}, \quad\; C_{(47,127)}=\{47,  94,  61, 122, 117, 107,  87\}, \\
&C_{(19,127)}=\{19 , 38,  76,  25,  50, 100,  73\}.
\end{aligned}
\end{equation*}
Thus $C_0^{(6,127)}=C_{(1,127)}\cup C_{(19,127)}\cup C_{(47,127)}$, which is equivalent to $g_0^{(6,127)}=M_{1}(x)M_{19}(x)M_{47}(x)$. And in this way, the following equations can be drawn easily.
\begin{equation*}
\begin{aligned}
&C_1^{(6,127)}=C_{(3,127)}\cup C_{(7,127)} \cup C_{(23,127)},\;C_2^{(6,127)}=C_{(9,127)}\cup C_{(11,127)} \cup C_{(21,127)}, \\
&C_3^{(6,127)}=C_{(5,127)}\cup C_{(27,127)} \cup C_{(63,127)}, \;C_4^{(6,127)}=C_{(13,127)}\cup C_{(15,127)} \cup C_{(31,127)}, \\
&C_5^{(6,127)}=C_{(29,127)}\cup C_{(43,127)} \cup C_{(55,127)}.
\end{aligned}
\end{equation*}
\end{example}
It is easy to deduce that the augmented cyclic code $C$ of $C_{i}$ (or $D_{i}$) can be obtained by removing one or more irreducible factors of $g_{i}^{(6,n)}$ (or $g_{i}^{(6,n)}(x)g_{i+1}^{(6,n)}(x)$ $g_{i+2}^{(6,n)}(x)$). It is also notable that any augmented code $C$ obtained by this way is also dual-containing. Furthermore, we have the following results.

\begin{lemma}\label{ACC}
Let $n=12m+7$ be an odd prime, where $m$ is a natural number. And $C_{i}$, $D_{i}$ be the cyclic codes defined above for any $i\in\{0,1,2,3,4,5\}$. Then, the following conclusions hold.

$(a)$ If $C=\langle \frac{g_{i}^{(6,n)}}{\prod_{i\in A}M_i(x)}\rangle$, then $C_i \subset C$, where $A$ is some proper subset of $\{i_1,i_2,...,i_t\}$.

$(b)$ If $D=\langle \frac{g_{i}^{(6,n)}(x)g_{i+1}^{(6,n)}(x)g_{i+2}^{(6,n)}(x)}{\prod_{i\in B}M_i(x)}\rangle$, then $D_i \subset D$, where $B$ is some proper subset of $\{i_1,i_2,...,i_t\} \cup \{(i+1)_1,(i+1)_2,...,(i+1)_t\} \cup \{(i+2)_1,(i+2)_2,...,(i+2)_t\}$.
\end{lemma}
$\mathbf{Proof.}$
The results come from the definition of dual-containing code. $\blacksquare$

\section{Quantum synchronizable codes from the cyclic codes obtained}\label{section5}
\hspace{1.5em}According to Lemma \ref{Theorem1}, our main goal is to find a pair of cyclic codes $C_1$ and $C_2$ of the same length and dimensions $k_1>k_2>\left \lceil \frac{n}{2} \right \rceil$, and satisfies the chain condition $C_2^{\perp}\subseteq C_2 \subset C_1$. In this section, we give construct a class of quantum synchronizable codes by the cyclic codes obtained above. To get a good quantum synchronizable code, the cyclic codes $C_1$ and $C_2$ are optimal or almost optimal. And analyzing the error-correcting capability of these quantum synchronizable codes.

\subsection{Maximum misalignment tolerance}
\hspace{1.5em}In the context of quantum synchronizable codes, we would like to maximize $\mathrm{ord}(f(x))$, where $f(x)$ is the quotient in Lemma \ref{Theorem1}. As $f(x)|(x^n-1)$, we know that the $\mathrm{ord}(f(x))$ is some divisor of $n$. Since $n=12m+7$ is an odd prime, $\mathrm{ord}(f(x))$ must be $n$. So, the tolerance capability of quantum synchronizable codes constructed is upper bounded by its length $n$.
We now give our main theorem.

Based on the cyclic code $C_i$ we constructed in Lemma \ref{lemma2}, we can obtain a class of quantum synchronizable codes as follows, whose synchronization capabilities attain the upper bound.
\begin{theorem}\label{QSC1}
Let $n=12m+7$ be an odd prime, $t=\frac{n-1}{6\ell} \geq 2k+1$ ($k$ is a positive integer) and $q\in C_0^{(6,n)}$, where $q^{\ell}\equiv 1 \bmod n$. For any nonnegative integers $c_l$ and $c_r$ satisfying $c_l+c_y <n$, then there exist a quantum synchronizable code with parameters $(c_l,c_r)-[[n+c_l+c_r,2z\ell+\frac{2n+1}{3}]]_q$, where $z$ is a nonnegative integer such that $0 \leq z\leq t-2 =\frac{n-12\ell-1}{6\ell}$.
\end{theorem}
$\mathbf{Proof.}$
Since $n=12m+7$ is an odd prime, then $|C_i^{(6,n)}|=\frac{n-1}{6}=2m+1$. Due to $|C_{(s,n)}|=\ell$ for any $s\in\{1,2,...,n-1\}$, and $\ell|(2m+1)$, it is obvious that cyclic code $C_i=\langle g_i^{(6,n)}(x)\rangle$ has augmented codes if and only if $g_i^{(6,n)}(x)$ has at least $2k+1$ irreducible factors over $\mathrm{F}_q$, where $k$ is a positive integer. So we let $t=\frac{n-1}{6\ell} \geq 2k+1$. According to $(a)$ in Lemma \ref{ACC}, cyclic code $C_{M_i}=\langle M_i(x)\rangle$ has a augmented code $C=\left[n,n- \mathrm{ord}(M_i(x))\right]=\left[n,n-\left( \frac{n-1}{6}-|A|\ell \right)\right]=\left[n,\frac{5n+1}{6}+|A|\ell\right]$, where $|A|$ is the size of the set $A$. Taking a set $A'$ such that $A\subset A' \subset \{i_1,i_2,...,i_t\}$, then we can get cyclic code $C'=\left[n,\frac{5n+1}{6}+|A'|\ell\right]$ and $C\subset C'$. Let $|A|=z$, then $0\leq z\leq t-2 =\frac{n-12\ell-1}{6\ell}$. Furthermore, we can obtain quantum synchronizable code with parameters $(c_l,c_r)-[[n+c_l+c_r,2z\ell+\frac{2n+1}{3}]]_q$ by Lemma \ref{Theorem1}, where $c_l+c_r <n$. $\blacksquare$

Moreover, we can also construct the other class of quantum synchronizable codes whose synchronization capabilities reach the upper bound, by using cyclic codes $D_i$ and their augmented codes.
\begin{theorem}
Let $n=12m+7$ be an odd prime, $t=\frac{n-1}{6\ell} \geq 2k+1$ ($k$ is a positive integer) and $q\in C_0^{(6,n)}$, where $q^{\ell}\equiv 1 \bmod n$. For any nonnegative integers $c_l$ and $c_r$ satisfying $c_l+c_y <n$, then there exist a quantum synchronizable code with parameters $(c_l,c_r)-[[n+c_l+c_r,2z\ell+1]]_q$, where $z$ is a nonnegative integer such that $0 \leq z\leq t-2 =\frac{n-12\ell-1}{6\ell}$.
\end{theorem}
$\mathbf{Proof.}$
Since $n=12m+7$ is an odd prime, then $|C_i^{(6,n)}|=\frac{n-1}{6}=2m+1$. Due to $|C_{(s,n)}|=\ell$ for any $s\in\{1,2,...,n-1\}$, and $\ell|(2m+1)$, it is obvious that cyclic code $D_i=\langle g_i^{(6,n)}(x)g_{i+1}^{(6,n)}(x)g_{i+2}^{(6,n)}(x)\rangle$ has augmented codes if and only if $g_{i+j}^{(6,n)}(x)$ ($j\in \{0,1,2\}$) has at least $2k+1$ irreducible factors over $\mathrm{F}_q$, where $k$ is a positive integer. So we let $t=\frac{n-1}{6\ell} \geq 2k+1$. According to $(b)$ in Lemma \ref{ACC}, cyclic code $D_{M_i}=\langle M_i(x)\rangle$ has a augmented code $D=\left[n,n- \mathrm{ord}(M_i(x))\right]=\left[n,n-\left( \frac{n-1}{2}-|B|\ell \right)\right]=\left[n,\frac{n+1}{2}+|B|\ell\right]$, where $|B|$ is the size of the set $B$. Taking a set $B'$ such that $B\subset B' \subset ( \{i_1,i_2,...,i_t\}\cup \{(i+1)_1,(i+1)_2$ $,...,(i+1)_t\} \cup \{(i+2)_1,(i+2)_2,...,(i+2)_t \} )$, then we can get cyclic code $D'=\left[n,\frac{n+1}{2}+|B'|\ell\right]$ and $D\subset D'$. Let $|B|=z$, then $0\leq z\leq t-2 =\frac{n-12\ell-1}{6\ell}$. Furthermore, we can obtain quantum synchronizable code with parameters $(c_l,c_r)-[[n+c_l+c_r,2z\ell+1]]_q$ by Lemma \ref{Theorem1}, where $c_l+c_r <n$. $\blacksquare$

\begin{example}
Let $n=127$ and $q=2\in C_{0}^{(6,n)}$. For convenience, we only consider the construction of quantum synchronizable codes from the cyclic code $C_1=\langle g_1^{(6,127)}(x) \rangle $ and its augmented cyclic code. In this case, $0\leq z\leq 1$, by Theorem \ref{QSC1}.

If $z=0$, then $C=C_1$ with parameters $[127,106]_2$ and $C^{\perp} \subset C$. If $z=1$, we can obtain $C'=\langle \frac{g_1^{(6,n)}(x)}{M_i(x)}\rangle$ with parameters $[127,113]_2$ and ${C'}^{\perp} \subset C'$, where $i\in\{3,7,23\}$. Furthermore, let $C''=\langle \frac{g_1^{(6,127)}(x)}{M_i(x)M_j(x)} \rangle$ with parameters $[127,120]_2$ and ${C''}^{\perp}\subset C''$ where $i\neq j \in \{3,7,23\}$. Then it is obvious that $C \subset C' \subset C''$. By taking $ C\subset C'$ and $C^{\perp} \subset C$, we have quantum synchronizable code with parameters $(c_l,c_r)-[[127+c_l+c_r,85]]_2$ and $c_l+c_r <127$,according to Lemma \ref{Theorem1}. And taking $C' \subset C''$ and ${C'}^{\perp} \subset C'$, we obtain quantum synchronizable code with parameters $(c_l,c_r)-[[127+c_l+c_r,99]]$ and $c_l+c_r < 127$.
\end{example}
\section{Conclusion}\label{section6}
\hspace{1.5em}We studied two classes quantum synchronizable codes from dual-containing cyclic codes obtained by sextic cyclotomic classes. Since $n=12m+7$ is an odd prime, these quantum synchronizable codes possess the highest possible tolerance against misalignment errors. By choosing some optimal dual-containing cyclic codes, the quantum synchronizable codes we get possess good error-correcting capabilities towards bit errors and phase errors.

\bibliography{Quantum_Synchronizable_Codes_on_Sextic_Cyclotomy}
\end{document}